\documentclass[twocolumn]{aastex631}

\usepackage{multirow}
\usepackage{comment}
\shorttitle{MAXI~J1803$-$298}
\shortauthors{Coughenour et al.}
\graphicspath{{./}{figures/}}
\begin{document}

\title{Reflection and timing study of the transient black hole X-ray binary MAXI~J1803$-$298 with NuSTAR}

%\title{Template \aastex Article with Examples: 
%v6.31\footnote{Released on March, 1st, 2021}}

% AUTHOR LIST
\author[0000-0003-0870-6465]{Benjamin M. Coughenour} \thanks{coughenour@berkeley.edu}
\affiliation{Space Sciences Laboratory, University of California, Berkeley,
7 Gauss Way, Berkeley, CA  94720-7450, USA}

\author[0000-0001-5506-9855]{John A. Tomsick}
\affiliation{Space Sciences Laboratory, University of California, Berkeley,
7 Gauss Way, Berkeley, CA  94720-7450, USA}

\author[0000-0003-4216-7936]{Guglielmo Mastroserio}
\affiliation{Cahill Center for Astronomy and Astrophysics, California Institute of Technology,
1216 E California Blvd, Pasadena, CA 91125, USA}

\author[0000-0002-5872-6061]{James F. Steiner}
\affiliation{Center for Astrophysics, Harvard \& Smithsonian,
60 Garden St, Cambridge, MA 02138, USA}

\author[0000-0002-8908-759X]{Riley M. T. Connors}
\affiliation{Department of Physics, Villanova University,
800 E Lancaster Avenue, Villanova, PA 19085, USA}
%\affiliation{Cahill Center for Astronomy and Astrophysics, California Institute of Technology,
%1216 E California Blvd, Pasadena, CA 91125, USA}

\author[0000-0002-9639-4352]{Jiachen Jiang}
\affiliation{Institute of Astronomy, University of Cambridge,
Madingley Road, Cambridge CB3 0HA, UK}

\author[0000-0002-8548-482X]{Jeremy Hare} \thanks{NASA Postdoctoral Program Fellow}
\affiliation{NASA Goddard Space Flight Center, Greenbelt, MD 20771, USA}

\author[0000-0002-8808-520X]{Aarran W. Shaw}
\affiliation{Department of Physics, University of Nevada, Reno, NV 89557, USA}

\author[0000-0002-8961-939X]{Renee M. Ludlam}
\affiliation{Department of Physics and Astronomy, Wayne State University,
666 W Hancock, Detroit, MI 48201, USA}

\author[0000-0002-9378-4072]{A. C. Fabian}
\affiliation{Institute of Astronomy, University of Cambridge,
Madingley Road, Cambridge CB3 0HA, UK}

\author[0000-0003-3828-2448]{Javier A. Garc\'{i}a}
\affiliation{Cahill Center for Astronomy and Astrophysics, California Institute of Technology,
1216 E California Blvd, Pasadena, CA 91125, USA}

\author[0000-0001-7532-8359]{Joel B. Coley}
\affiliation{Department of Physics and Astronomy, Howard University,
Washington, DC 20059, USA}
\affiliation{CRESST/Code 661 Astroparticle Physics Laboratory, NASA Goddard Space Flight Center,
Greenbelt Rd., MD 20771, USA}

%\author{Others}
%\affiliation{Several Universities and Research Institutions}

\begin{abstract}

The transient black hole X-ray binary MAXI~J1803$-$298 was discovered on 2021 May 1, as it went into outburst from a quiescent state. As the source rose in flux it showed periodic absorption dips and fit the timing and spectral characteristics of a hard state accreting black hole. We report on the results of a Target-of-Opportunity observation with NuSTAR obtained near the peak outburst flux beginning on 2021 May 13, after the source had transitioned into an intermediate state. MAXI~J1803$-$298 is variable across the observation, which we investigate by extracting spectral and timing products separately for different levels of flux throughout the observation. Our timing analysis reveals two distinct potential QPOs which are not harmonically related at $5.4\pm0.2$\,Hz and $9.4\pm0.3$\,Hz, present only during periods of lower flux. With clear relativistic reflection signatures detected in the source spectrum, we applied several different reflection models to the spectra of MAXI~J1803$-$298. Here we report our results, utilizing high density reflection models to constrain the disk geometry, and assess changes in the spectrum dependent on the source flux. With a standard broken power-law emissivity, we find a near-maximal spin for the black hole, and we are able to constrain the inclination of the accretion disk at $75\pm2$ degrees, which is expected for a source that has shown periodic absorption dips. We also significantly detect a narrow absorption feature at $6.91\pm0.06$\,keV with an equivalent width between 4 and 9\,eV, which we interpret as the signature of a disk wind.

\end{abstract}


\section{Introduction} \label{sec:intro}

There are dozens of known Galactic black hole (BH) X-ray binaries \citep[XRBs,][]{tetarenko16}, yet many more quiescent BH systems likely exist in the Milky Way and remain undetected. These transient systems are typically only discovered in an outburst, when a sudden and dramatic increase in the accretion rate onto the BH results in an increase in luminosity that scales several orders of magnitude, from $\lesssim\,10^{34}$\,erg\,s$^{-1}$ to $\gtrsim\,10^{38}$\,erg\,s$^{-1}$. Transient BH XRBs may vary from a fraction of a percent of the Eddington limit ($L_{\rm Edd}$) for spherical accretion onto a BH to near-Eddington luminosities. Over the course of an outburst, a BH XRB will move through several different spectral states, tracing out a common pattern on its hardness-intensity diagram \citep[HID, see][for a review]{rm06}. Systems almost always rise in the hard state, when the spectrum is dominated by emission modeled by a power law or by the Compton up-scattering of soft, thermal disk photons. Near the peak of the outburst, sources will transition into an intermediate state, in which a thermal disk component becomes apparent in addition to the hard power law. Eventually, the disk component will dominate the spectrum and the source will have moved to the soft state. The standard theory which explains this spectral evolution over the course of an outburst is that the accretion disk is initially truncated, but moves toward the innermost stable circular orbit (ISCO) during the intermediate and soft state, providing the soft, thermal spectrum \citep{homan05, belloni05}.

%%% FIGURE 1: fig:lchc
\begin{figure*}[t]
\begin{center}
    \includegraphics[width=11cm,trim=1.4cm 1.4cm 2cm 2cm, clip]{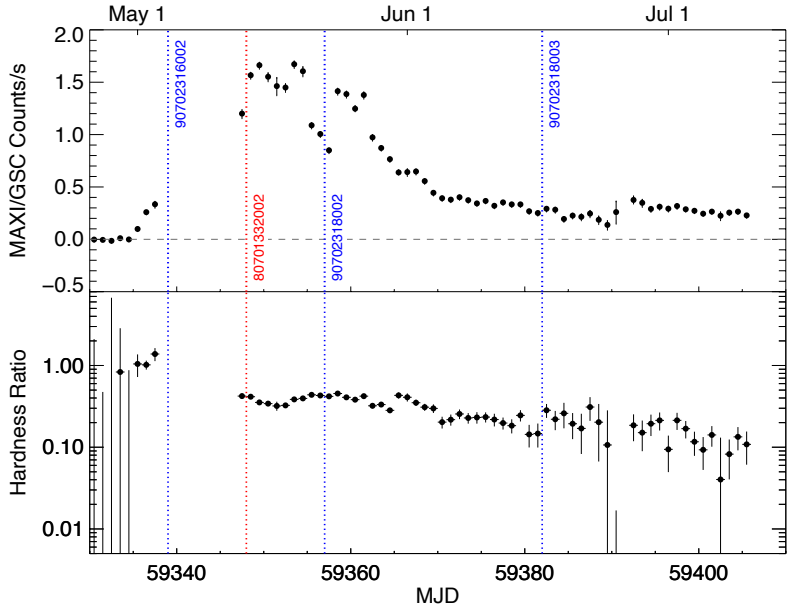}
    \caption{The full MAXI/GSC 2--20\,keV light curve (top) illustrates the source outburst over several months, with red and blue vertical dashed lines representing the four NuSTAR observations taken throughout the source's main activity. We also plot the MAXI/GSC hard color ratio, defined as the 4--20\,keV count rate divided by the 2--4\,keV count rate (bottom). Of the NuSTAR observations shown, this work is focused on the observation with the highest count rate, shown in red.}
\label{fig:lchc}
\end{center}
\end{figure*}

While there is some debate over the geometry of the accretion disk during the hard state, namely whether or not it extends to the ISCO \citep[see, e.g.][]{kolehmainen11, garcia15, miller15, basak16, zdziarski21, connors22}, it is generally accepted that the accretion disk arrives at the ISCO between the intermediate and soft states. The intermediate state is particularly interesting because both the soft, thermal disk and the hard power law component are quite evident in the spectrum. The interaction between the hard X-rays and the disk produces fluorescent and reprocessed emission, referred to as reflection. Reflection appears as a number of distinct features above the continuum, the most notable being a broad, asymmetric Fe K$\alpha$ line between 6 and 7\,keV \citep{fabian89, laor91, lightman88}. The `Compton hump' above $\sim$15\,keV is another distinct feature of reflection. The profile of the reflection spectrum, and in particular the Fe K$\alpha$ line, is shaped by the relativistic effects experienced by material in the innermost accretion disk, and as a result the reflection spectrum provides a way to measure details of the accretion environment in the immediate vicinity of the BH \citep{fabian89, fabian00, miller07}.

One of the key properties of a BH is its spin, which may be measured in the case of XRBs using reflection spectroscopy. When a BH rotates, the ISCO is located nearer to the event horizon than is the case for a non-rotating BH. When a source is in the intermediate or soft state and the accretion disk extends down to the ISCO, measurements of the BH spin (due to its effect on the location of the ISCO) are possible. With reflection, spin measurements of a number of BHs have been made (e.g., Cyg~X-1, \citealt{tomsick14, parker15, walton16}, GX~339$-$4 \citealt{parker16}, V404~Cyg \citealt{walton17}, XTE~J1908+094 \citealt{draghis21}; see \citealt{reynolds21} for a review). When the distance, BH mass, and inclination of a source are known, modeling the disk continuum also provides a measure of the ISCO, and may also constrain the BH spin \citep[see][for a review]{mcclintock14}.

Accretion disk densities of $n_e \sim 10^{15}$\,cm$^{-3}$ are to be expected for the more massive active galactic nuclei (AGN), while the smaller physical scales of stellar-mass BHs should produce disk densities well above $10^{15}$\,cm$^{-3}$ \citep{sz94}. The effects of higher disk densities in stellar-mass BH systems has only recently been explored with updated reflection models \citep{garcia16, tomsick18, jiang19a, connors21}, as well as with lower-mass AGN \citep{jiang19b, mallick22}. One prominent effect is that high-density models reduce the need for extreme super-solar abundances when fitting the reflection spectrum in XRBs, as was demonstrated in the case of Cygnus X-1 by \cite{tomsick18}. Another important effect, particularly with respect to the shape of the spectrum, is additional flux at low energies due to an increased opacity from free-free absorption that arises at higher disk densities \citep{rf07, garcia16}, though this effect is not obvious in the NuSTAR band ($> 3$\,keV) for densities below $n_e \sim 10^{20}$\,cm$^{-3}$ \citep{tomsick18, jiang22}.

The recently discovered transient X-ray binary MAXI~J1803$-$298 (hereafter MAXI~J1803) went into outburst on 2021 May 1 and was initially detected by MAXI/GSC \citep{serino21, shidatsu22}. Follow-up observation in the X-ray band alongside optical SALT spectroscopy suggested that the source is an accreting BH rising in the hard state \citep{bult21, buckley21, homan21, xu21}. Early in its outburst, while MAXI~J1803 was still in the hard state and before it reached its outburst peak, periodic absorption dips with a $\sim7$ hour cadence were detected with NICER, NuSTAR, and AstroSAT \citep{homan21, xu21, jana22}. Sometime after 2021 May 4, during an 8 day period in which the source was outside the MAXI field of view, it transitioned from the hard state into an intermediate state \citep{shidatsu21}, and then continued to increase in flux until reaching the peak of its outburst on 2021 May 16. On shorter timescales, quasi-periodic oscillations (QPOs) were seen in MAXI~J1803 with NICER, NuSTAR, Insight-HXMT, and AstroSat, ranging from 0.13\,Hz in the early hard state to 7.61\,Hz after the state transition \citep{bult21, xu21, wang21, chand21, chand22, jana22, ubach21}. Other noteworthy investigations of MAXI~J1803 report disk wind signatures in the optical and X-ray bands \citep{miller21, matasanchez22}, while recent X-ray spectral studies suggest MAXI~J1803 contains a rapidly spinning BH \citep{chand22, feng22}.

In this paper, we report the results of a Target-of-Opportunity (ToO) observation of MAXI~J1803 with NuSTAR, beginning on 2021 May 14 when the source was near its peak outburst flux. In Section~\ref{sec:data} we discuss the observation itself and data reduction, and in Section~\ref{sec:timing} we present our timing analysis. In Section~\ref{sec:spectra} we describe our spectral modeling and results. In Sections~\ref{sec:discuss} and \ref{sec:conclude} we consider the implications of these results for MAXI~J1803 as well as in the context of other studies of accreting BH XRBs.

\section{Observation \& Data Reduction} \label{sec:data}

%%% FIGURE 2: fig:lchid
\begin{figure*}[ht]
\begin{center}
    \includegraphics[width=8.5cm,trim=2.5cm 1.8cm 2cm 2.5cm, clip]{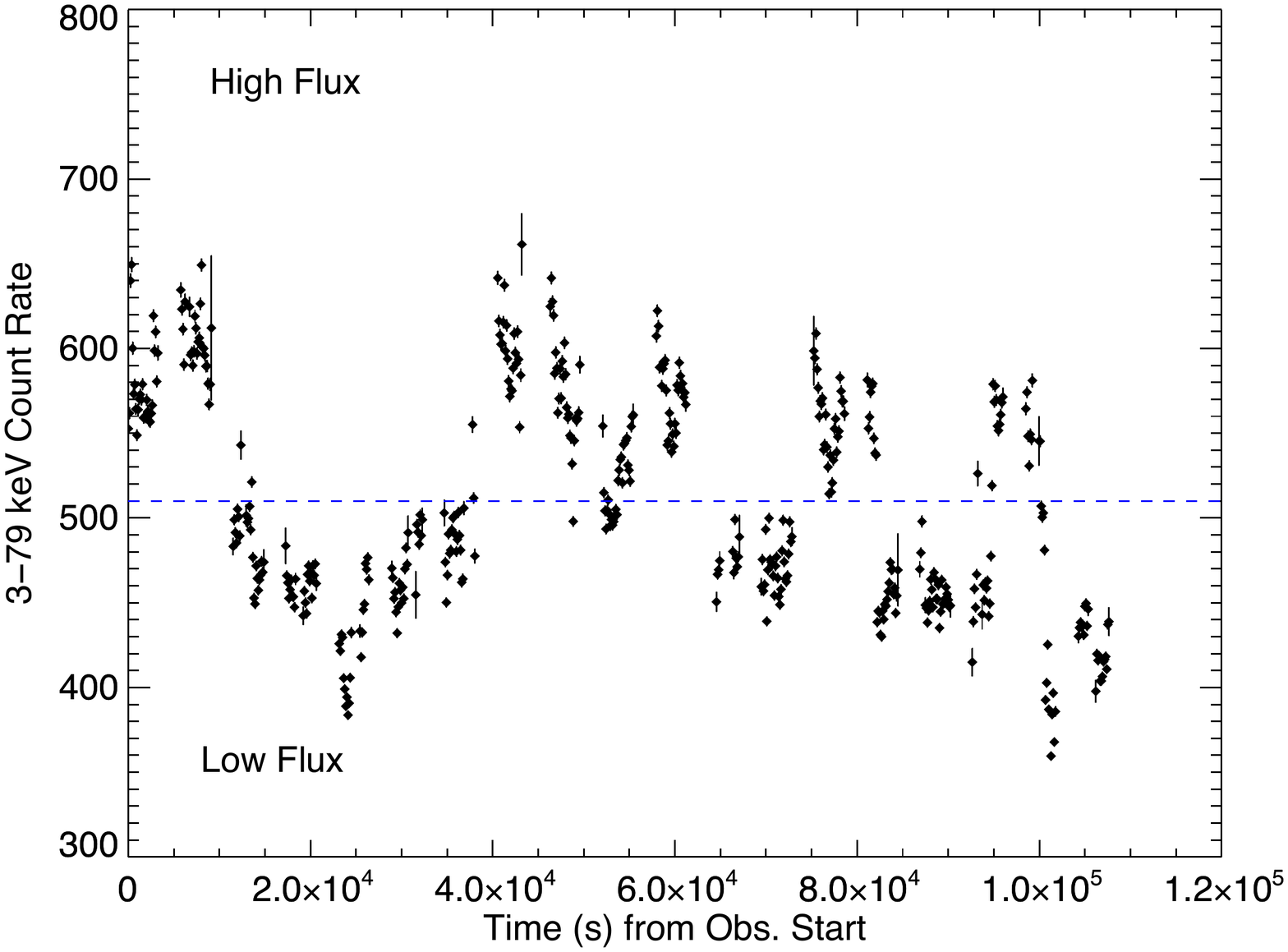}
    \includegraphics[width=8.5cm,trim=2.5cm 1.8cm 2cm 2.5cm, clip]{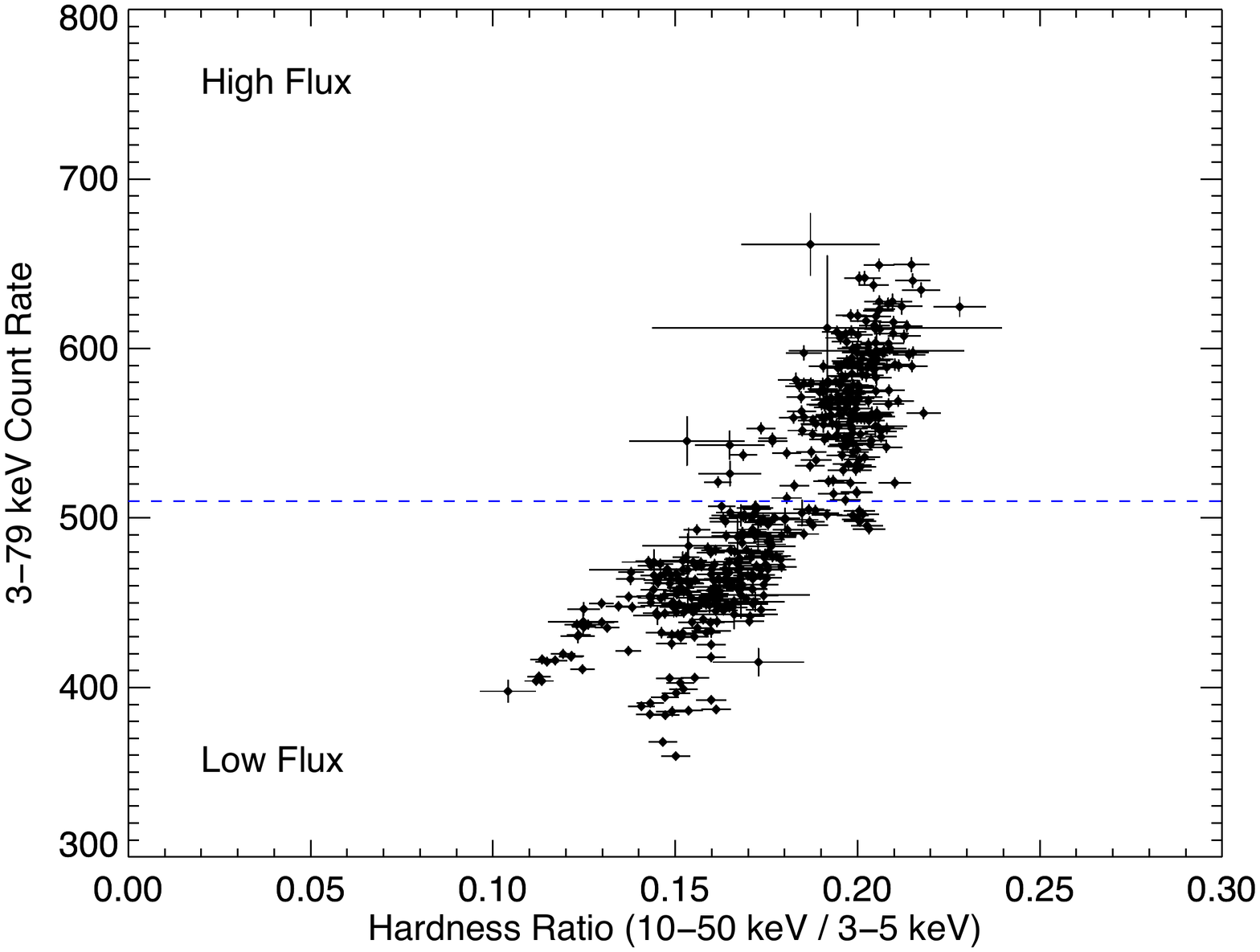}
    \caption{NuSTAR light curve (FPMA$+$B, left) and HID (right), in 128\,s bins and with the hard color taken as the ratio of the 10--50\,keV count rate to the 3--5\,keV count rate. Throughout our NuSTAR observation, MAXI~J1803 appears to switch between two distinct fluxes. We include a dashed line at 510 counts~s$^{-1}$ which we use to separate the observation into high and low flux components. NuSTAR Obsid 80701332002 began at 2021-05-14 23:01:09.}
\label{fig:lchid}
\end{center}
\end{figure*}

MAXI~J1803 was observed with NuSTAR several times throughout its 2021 outburst, the longest of which was a ToO observation near the peak of the outburst (see Figure~\ref{fig:lchc}). That observation (ObsID 80701332002) is the focus of this work, and started at 23.02 hours (UTC) on May 14, continuing until 05.35 hours on May 16, resulting in over 30\,ks of exposure in each Focal Plane Module (respectively called FPMA and FPMB) after dead time corrections and Earth occultations.

The data were reduced using NUSTARDAS~v2.1.1 (with HEASoft~v6.29) and CALDB 20211103, with the parameter `\texttt{statusexpr}' set to  \texttt{"STATUS==b0000xxx00xxxx000"}, which is recommended for bright sources\footnote{https://heasarc.gsfc.nasa.gov/docs/nustar/nustar\_faq.html}. Event lists and images were produced with \texttt{nupipeline}. On the order of hundreds of counts per second in each FPM, the source MAXI~J1803 dominates the images, though there was enough room to select source and background regions of 120'' radius on the same detector chip. Source and background spectra were then extracted using \texttt{nuproducts}, as well as light curves in 128\,s bins across a variety of energy bands to produce a hardness-intensity diagram (HID) alongside the full 3--79\,keV NuSTAR light curve. For the hard-color ratio, we divided the 10--50\,keV count rate by the 3--5\,keV count rate, avoiding the Fe line region to be sure that the continuum, rather than reflection spectral features, drive any changes in the source hardness. Spectra were grouped to a minimum of 30 counts per bin, and MAXI~J1803 dominates the background across the entire NuSTAR energy band, so we use the full 3--79\,keV for spectral analysis. The HID is plotted next to the broadband NuSTAR light curve in Figure~\ref{fig:lchid}.

Some variability is clear in the NuSTAR light curve, and MAXI~J1803 appears to occupy one of two distinct positions on its HID (Figure~\ref{fig:lchid}). To investigate this variability, and what spectral components are driving it, we extracted separate spectra for time intervals when MAXI~J1803 has greater than or fewer than 510 counts per second (with respect to the sum of both FPMs, this division is shown as the dashed line in Figure~\ref{fig:lchid}). Good time interval (GTI) files were produced using \texttt{dmgti} from the Chandra data analysis software package, CIAO v4.13, and then called using \texttt{nuproducts} to extract spectra. Throughout the paper, we refer to these as the `high' and `low' flux spectra, respectively.

\section{Timing Analysis} \label{sec:timing}

\begin{figure*}
\begin{center}
    \includegraphics[width=8.5cm,trim=0cm 0cm 0cm 0cm, clip]{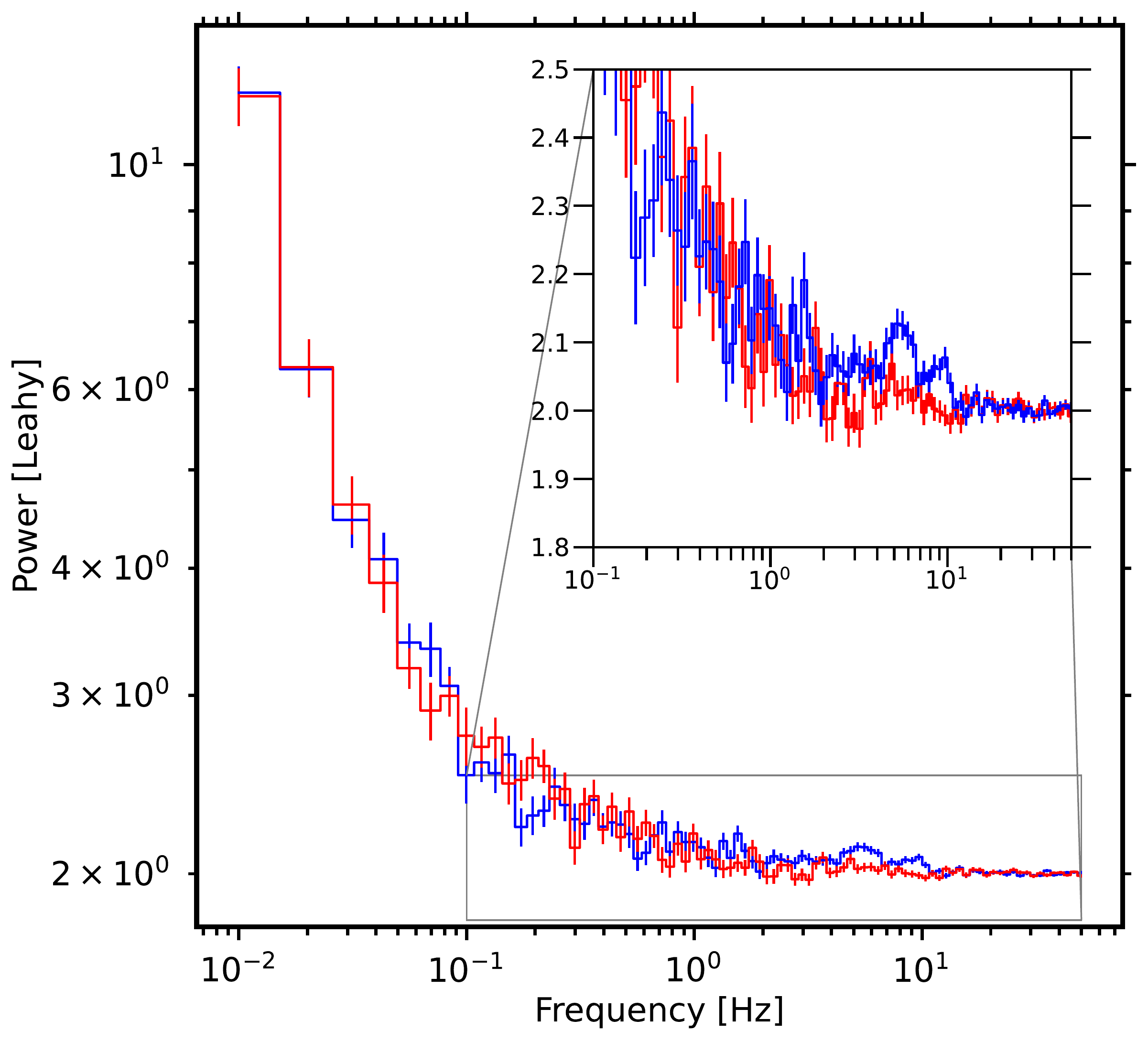}
    \raisebox{0.06\height}{\includegraphics[width=8.5cm,trim=2cm 1.5cm 2cm 1.5cm, clip]{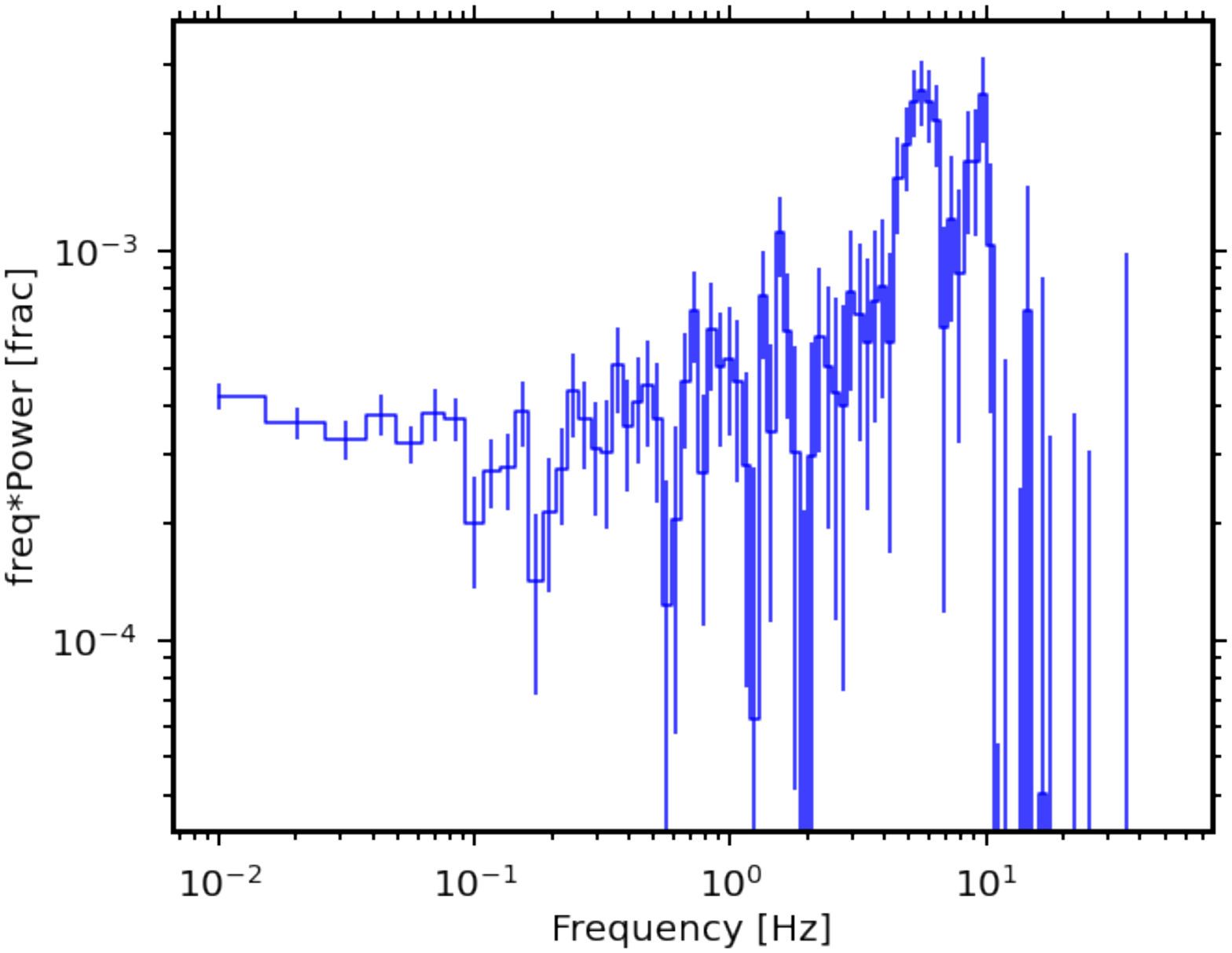}}
    \caption{Leahy power spectrum (left), calculated separately for the high flux (red) and low flux (blue) time intervals (see Figure~\ref{fig:lchid}), with an inset to highlight the two potential QPOs at 5.4\,Hz and 9.4\,Hz present in the low flux (blue) power spectrum. The right panel shows just the low flux PDS, scaled with frequency to illustrate the two potential QPOs rather than the low-frequency red noise.}
\label{fig:pds}
\end{center}
\end{figure*}

During this observation MAXI~J1803 shows limited variability over short ($\lesssim 100$\,s) timescales. The power density spectra (PDS) for the high and low flux lightcurves were calculated separately from 0.01\,Hz to 50\,Hz, and these are shown in Figure~\ref{fig:pds}. The PDS was computed using the Stingray software \citep{huppenkothen19a, huppenkothen19b}, which allows us to apply a dead time correction comparing the two modules \citep{bachetti15}. Mostly featureless, the PDS is dominated by red noise at low frequencies and Poisson noise at higher frequencies. It does not show a strong QPO like the previous NuSTAR observation on May 5 (at 0.41\,Hz), or any broadband noise component at higher frequencies. Prior to our May 14 observation, a Type~C QPO was seen on in MAXI~J1803 on May 12 with AstroSat, with a frequency that evolved from 5.31\,Hz to 7.61\,Hz over the course of that observation \citep{jana22}. Along with that QPO, as is expected for Type~C QPOs, was a flat broadband noise component in the PDS. Just days later, the PDS of the source is markedly different, but not entirely featureless - two weak yet significant bumps are visible in the low flux PDS just below $\sim$10\,Hz, and these are highlighted in the Figure~\ref{fig:pds} inset. Curiously, no features are seen in the higher flux PDS, when the source spectrum is slightly harder.

%%%%% FIGURE 3 %%%%%

%%%%% TABLE 1 %%%%%
\begin{table}
\begin{minipage}{\columnwidth}
\centering
\caption{Power Spectrum Modeling Results \label{tab:lorentz}}
\renewcommand{\arraystretch}{1.25}
\vspace{-0.3cm}
\begin{tabular}{lccc}
\hline \hline
Model & Parameter & Units & Value \\ \hline
Poisson Noise & Constant & --- & $2.000 \pm0.003$ \\
\multirow{2}{*}{Power Law} & Norm. & --- & $0.080 \pm0.009$ \\
  & Index & --- & $1.02 \pm0.03$ \\
\multirow{4}{*}{Lorentzian$_{1}$} & Norm$_1$ & --- & $0.102^{+0.015}_{-0.013}$ \\
  & $\nu_{1}$ & Hz & $5.4 \pm0.2$ \\
  & $\Delta_1$ & Hz & $1.6 \pm0.4$ \\
  & $Q_1$ & $\nu_1 / (2\,\Delta_1)$ & 1.8 \\
\multirow{4}{*}{Lorentzian$_{2}$}   & Norm$_2$ & --- & $0.056^{+0.020}_{-0.018}$ \\
  & $\nu_{2}$ & Hz & $9.4 \pm0.3$ \\
  & $\Delta_2$ & Hz & $0.8^{+0.4}_{-0.3}$ \\ 
  & $Q_2$ & $\nu_2 / (2\,\Delta_2)$ & 6.1 \\ \hline
\end{tabular} \\
\flushleft{\footnotesize With the Leahy normalization, a constant of 2 represents Poisson noise. We also apply a power law to fit the low frequency red noise component, and two Lorentzians for the possible QPOs. Each Lorentzian function peaks at its centroid frequency $v_{1,2}$ with a half-width half-maximum of $\Delta_{1,2}$. Errors shown represent 95\% confidence intervals.}
\end{minipage}
\end{table}

We tested for the significance of these possible QPO bumps by fitting the PDS with a model combining a constant, a power law, and two Lorentzians, to represent the Poisson noise, red noise, and the two bumps, respectively. The fit was performed with both maximum likelihood statistics and Bayesian parameter estimation assuming flat priors on every parameter, using the Stingray internal routines. The parameter values are compatible between the two methods, and we report the Bayesian results and 95\% confidence intervals in Table~\ref{tab:lorentz}. First, our results show that the two potential QPO signals are not harmonically related, with $\nu_1 = 5.4 \pm0.2$\,Hz and $\nu_2 = 9.4 \pm0.3$\,Hz (quoting 95\% confidence intervals). The bumps are broad, with full-width half-maxima of $2\,\Delta_1 = 3.2^{+0.7}_{-0.8}$\,Hz and $2\,\Delta_2 = 1.5^{+0.8}_{-0.6}$\,Hz, corresponding to quality-factors of $Q_1 = 1.8$ and $Q_2 = 6.1$, respectively. The bumps are therefore broad enough to hardly be considered QPOs (particularly the one at 5.4\,Hz). Both are significant, however, with a non-zero Lorentzian normalization preferred by the model at almost $12\,\sigma$ for the lower frequency component and at almost $5\,\sigma$ for the higher frequency component. We also computed the fractional rms in the frequency range 0.01-10 Hz for both high and low flux light curves --- the values are 6$\%\pm2\%$ and 9$\%\pm2\%$, respectively. These values are dominated by the low frequency red noise, which is the strongest component in the PDS. The long timescale variability that causes the source to oscillate between periods of high and low flux (see Fig.~\ref{fig:lchid}) is not considered in the rms calculation, because those variations appear at longer timescales than 100 seconds. In order to investigate changes in the source on these timescales, we continue with our spectral analysis in the next section, and we consider the phenomenology of the potential QPOs in the Discussion.

\section{Spectral Analysis} \label{sec:spectra}

%%%%% FIGURE 4: FPMA/B %%%%%
\begin{figure}
\begin{center}
    \includegraphics[width=8.5cm,trim=1.2cm 3.5cm 2.4cm 5.2cm, clip]{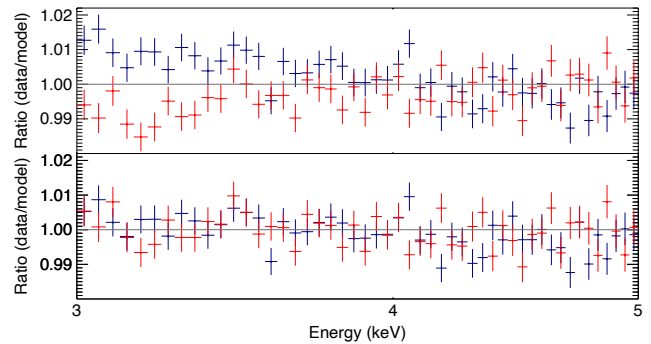}
    \caption{Disagreement between FPMA (navy) and B (red) below 4\,keV, shown as the normalized residuals of the data to a reflection model with $N_{\rm H} = 0.32$ for both FPMs (top), and with the neutral column density allowed to vary for FPMB (bottom). This results in an artificially high $N_{\rm H}$ for FPMB but prevents the difference between the two FPMs from affecting our spectral fit results.}
\label{fig:nH}
\end{center}
\end{figure}

%%%%% FIGURE 5: Reflection %%%%%
\begin{figure}
\begin{center}
   \includegraphics[width=8.5cm,trim=2.5cm 2cm 3cm 2.5cm, clip]{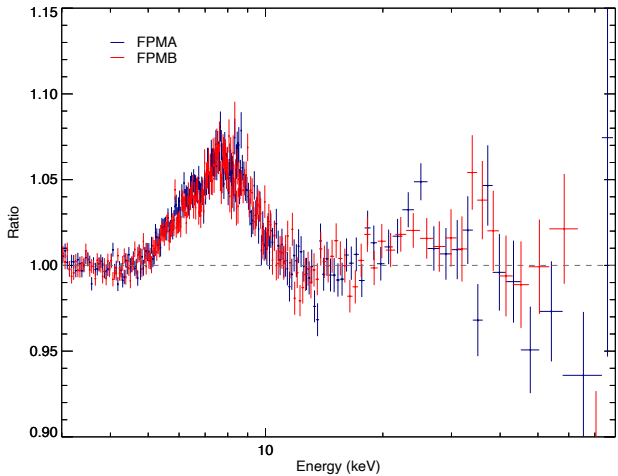}
    \caption{Ratio of the full, time-averaged NuSTAR spectrum to a model, consisting of an absorbed multi-temperature disk and power law, with energy bands ignored to highlight the reflection features. The most notable signature of reflection is the extremely broad Fe line, centered between 6 and 7\,keV, though the Compton hump above $\sim$20\,keV is also noticeable.}
\label{fig:feline}
\end{center}
\end{figure}

Spectra were modeled using XSPEC v12.12.0 \citep{arnaud96} with $\chi^2$ statistics. To simultaneously fit spectra from FPMA and FPMB, a constant multiplicative factor was included in each model, set to 1.0 for FPMA and allowed to vary freely for FPMB.

An unusual discrepancy between the two FPMs was visible in the spectra at $\sim$3\,keV, with FPMA systematically higher than FPMB, while the two spectra are in good agreement above 4\,keV. Although this is similar in effect to a difference in temperature between the two FPMs due to a tear in the multi-layer insulation (MLI) covering FPMA \citep[see][]{madsen20}, the difference between the two FPMs is the result of gain corrections from the updated CALDB 20211103 \citep{madsen21}, and does not occur when the data are processed with previous CALDB versions. The detector response is calculated across each FPM without taking into account pixel-to-pixel variations, which is responsible for the small differences ($\lesssim$2\%) seen in the MAXI~J1803 spectra. In order to prevent this discrepancy from affecting the spectral modeling, we allowed for the neutral absorption column density of FPMB to be different than that of FPMA, and though unphysical this brought the two spectral models into good agreement, as shown in Figure~\ref{fig:nH}. For FPMA, the column density was fixed to $N_{\rm H} = 3.2\times10^{21}$\,cm$^{-2}$, based on the reported NICER best-fit values for MAXI~J1803, and considering that NICER's soft X-ray coverage will better constrain absorption than NuSTAR \citep{bult21,homan21}. The value for FPMB, meanwhile, was free to vary (with best-fit values of $\sim 6.0\times10^{21}$\,cm$^{-2}$, regardless of the continuum model). We checked throughout our analysis that this choice of column densities for FPMA/B does not impact our results (as compared to fixing $N_{\rm H}$ for FPMB). Neutral absorption along the line of sight was modeled using \textsc{tbabs}, with Wilms abundances \citep{wilms00} and Verner cross-sections \citep{verner96}. We quote 90\% confidence limits on our spectral parameters, unless otherwise specified.

We began fitting the full time-averaged spectrum with a basic phenomenological model, namely a multi-temperature disk blackbody model (\textsc{diskbb} in XSPEC) as well as a simple power law (the data showed no sign of a high-energy cutoff, so \textsc{powerlaw} in XSPEC was used). This model provided a poor fit to the data, with $\chi^2$/degrees of freedom (dof) = 3300.7/2259, and large residuals between 5 and 10\,keV owing to the presence of broad Fe emission. Figure~\ref{fig:feline} illustrates this, showcasing the reflection features present in the spectrum when the disk and power law model is fit to the data while ignoring the Fe line region. When fitting the full spectrum, we measure the inner disk temperature to be $1.138\pm0.002$\,keV, and for the power law photon index we measure $\Gamma = 2.52\pm0.01$. Considering the variability inherent in our NuSTAR observation, we applied our basic phenomenological model to the high and low flux spectra (see Section~\ref{sec:data}) separately, and we found that the differences between the two spectra are primarily due to the \textsc{powerlaw} normalization. It is also worth mentioning that, while periodic absorption dips have been detected for this source \citep{homan21, xu21, jana22}, the low flux spectra cannot be explained by an increase in the absorption column relative to the high flux spectra. Furthermore, the observed changes in flux are not periodic (see Fig.~\ref{fig:lchid}).

Given the clear reflection features visible in the residuals of our phenomenological fits (see Figure~\ref{fig:feline}), we applied a series of reflection models in addition to the disk, beginning with the standard \textsc{relxill} model v1.4.3 \citep{garcia14, dauser14}. The \textsc{relxill} model takes into account reflection and reprocessing of hard X-rays by the accretion disk, as well as the relativistic effects that asymmetrically broaden distinct features such as the Fe~K-shell emission at 6.4\,keV. Aside from the broad Fe line, the Compton `hump' above ~20\,keV is an important feature that the high-energy coverage of NuSTAR is uniquely able to constrain. Considering the differences between the high and low flux spectra in our phenomenological fits, we fit the \textsc{diskbb}+\textsc{relxill} model to these spectra separately.

The \textsc{relxill} model allows for a broken power law disk emissivity, with two indices $q_{1}$ and $q_{2}$ each defining the emissivity profile of the accretion disk above and below a break radius, $R_{\rm break}$, respectively. We fixed the disk emissivity index to be $q=3$ throughout the disk, the inner radius to be 1\,ISCO (equal to $6\,GM/c^2 = 6\,R_g$ for a Schwarzchild BH, or just 1\,$R_g$ for a maximally spinning Kerr BH), and the outer disk radius to be 1000\,$R_g$. Because MAXI~J1803 exhibited absorption dips in the early phases of its outburst with a period of $\sim$7 hours \citep{homan21,xu21,jana22}, its orbit is highly inclined relative to our line of sight, and we expect the accretion disk inclination as measured by \textsc{relxill} to similarly be a high value, but we left it free to vary. Finally, just as the initial power law showed no signs of a cutoff in the NuSTAR band, the same is true of the power law incident flux in \textsc{relxill}, and we fix the energy cutoff parameter to 300\,keV. We then fit both the high and low flux spectra together, tying the BH spin and disk inclination between the two, and this model (a multi-temperature disk in addition to \textsc{relxill}) provided a much better fit to the data with $\chi^{2}$/dof = 4117.5/3770.

We also tested different versions of the standard \textsc{relxill} model, including \textsc{relxillCp} and \textsc{relxillLp}. While \textsc{relxillCp} and its variations consistently provided a much poorer fit to the data, our spectral fits were significantly improved for both the high and low flux spectra when using the \textsc{relxillLp} model. This version assumes the lamppost geometry where a point source some height $h$ above the BH provides the hard X-ray flux of the corona. The hard X-ray emission spectrum is a power law in both \textsc{relxill} and \textsc{relxillLp}. Under the lamppost geometry, the disk emissivity and reflection fraction can be calculated self-consistently --- this is done for \textsc{relxillLp} in v1.4.3 by setting the `\texttt{fixReflFrac}' parameter to 1. With the inner and outer disk radii fixed to 1\,ISCO and 1000\,$R_g$, respectively, the spin limited to $0 \le a \le 0.998$, and the Fe abundance fixed to solar values, our lamppost model provided a marginally acceptable fit ($\chi^{2}$/dof = 4077.9/3770) and an improvement over our results using \textsc{relxill} with a fixed disk emissivity. Using the lamppost model and allowing the Fe abundance to vary, we found $A_{\rm Fe} = 4.9^{+1.0}_{-0.3}$ times the solar value, with $\Delta\chi^2 = 55.8$, representing a significant improvement in the fit. While a high Fe abundance has been commonly seen in reflection studies of BH XRBs \citep{garcia18}, in many cases it has been shown that this is a limitation of reflection modeling rather than a true characterization of the accretion disk. Perhaps the most promising solution has been the development of high-density reflection models, which allow for greater densities within the disk than the standard value of $n_e$ = 10$^{15}$\,cm$^{-3}$, which is typical for active galactic nuclei but lower than would be expected for a stellar-mass BH \citep{garcia16}. High-density reflection models have provided better and more physically realistic constraints on Fe abundances \citep{tomsick18, jiang19a}, with $A_{\rm Fe} \lesssim 3$ times the solar value \citep[see, e.g.,][]{feltzing01, taylor03}.

We applied the high density versions of \textsc{relxill} and \textsc{relxillLp} to our MAXI~J1803 spectra, with the same constraints as before. In both cases, the high density models provided an improved fit, while the lamppost continued to out-perform the standard high density model \textsc{relxillD} as long as the disk emissivity was fixed. However, like its standard version, \textsc{relxillD} allows for a broken power law accretion disk emissivity. Alongside our results with the high density lamppost model, \textsc{relxillLpD}, we fit the spectra of MAXI~J1803 using \textsc{relxillD} with $1 \le q_{1} \le 10$, $R_{\rm break} \le 10\,R_g$, and $q_{2} = 3$. Allowing for a more complex inner disk emissivity profile significantly improves the fit, as does allowing for higher densities, and we find a fit statistic of $\chi^{2}$/dof = 4011.4/3767 using \textsc{relxillD}. This also marks an improvement over the high density lamppost, with $\chi^{2}$/dof = 4024.0/3768. Throughout the rest of the paper, we compare these two different models (utilizing either \textsc{relxillD} or \textsc{relxillLpD}) and their different interpretations.

%%%%% TABLE 2: Results with Lamppost & Gaussian, A_Fe free %%%%%
\begin{table*}
\begin{minipage}{2\columnwidth}
\caption{Co-Fit Spectral Results with Model 1 \label{tab:relxillLpDgauss}}
\centering
\renewcommand{\arraystretch}{1.25}
\begin{tabular}{lcccc}
\multicolumn{5}{c}{\textsc{constant$^\ast$tbabs$^\ast$(diskbb + relxillLpD + gaussian)}} \\
\hline \hline
Model & Parameter & Units & Low Flux Value & High Flux Value \\ \hline
\textsc{constant} & --- & FPMA/B &  \multicolumn{2}{c}{ $0.981 \pm0.002$ } \\ \hline
\multirow{2}{*}{\textsc{tbabs}}
  & $N_{\rm H}$ (A) & $10^{21}$\,cm$^{-2}$ & \multicolumn{2}{c}{ 3.2~$^{a}$ } \\
  & $N_{\rm H}$ (B) & $10^{21}$\,cm$^{-2}$ & \multicolumn{2}{c}{ $6.3 \pm0.4$ } \\ \hline
\multirow{2}{*}{\textsc{diskbb}}
  & $T_{\rm in}$ & keV & $1.064 \pm0.004$ & $1.137^{+0.005}_{-0.004}$ \\
  & $N$ & --- & $455 \pm8$ & $307 \pm7$ \\ \hline
\multirow{10}{*}{\textsc{relxillLpD}}
  & $h$ & $GM/c^2$ & $31^{+8}_{-7}$ & $45^{+18}_{-15}$ \\
  & $a$ & --- & \multicolumn{2}{c}{ $0.996^{~b}$ } \\
  & Incl. & deg. & \multicolumn{2}{c}{ $76^{+4}_{-7}$ } \\
  & $R_{\rm in}$ & ISCO & \multicolumn{2}{c}{ 1.0~$^{a}$ } \\
  & $R_{\rm out}$ & $GM/c^2$ & \multicolumn{2}{c}{ 1000~$^{a}$ } \\
  & $\Gamma$ & --- & $2.37 \pm0.02$ & $2.41 \pm0.03$ \\
  & log\,$\xi$ &  log\,(erg\,cm\,s$^{-1}$) & $3.68^{+0.07}_{-0.05}$ & $3.92^{+0.07}_{-0.09}$ \\
  & $A_{\rm Fe}$ & solar & \multicolumn{2}{c}{ $1.4^{+0.2}_{-0.4}$ } \\
  & log\,$n_e$ & log\,(cm$^{-3}$) & \multicolumn{2}{c}{ $> 18.76$ } \\
  & $f_{refl}$ & --- & $1.00^{~c}$ & $1.07^{~c}$ \\
  & $N$ & $10^{-2}$ & $2.5^{+0.3}_{-0.4}$ & $3.7 \pm0.6$ \\ \hline
  & $E_{\rm line}$ & keV & \multicolumn{2}{c}{ $6.91 \pm0.06$ } \\
\textsc{gaussian} & $\sigma_{\rm line}$ & eV & \multicolumn{2}{c}{ 50~$^{a}$ } \\
  & $N_{\rm line}$ & $10^{-4}$\,ph\,cm$^{-2}$\,s$^{-1}$ & \multicolumn{2}{c}{ $-3.6 \pm1.1$ } \\ \hline  
  & $\chi^{2}$/dof & --- & \multicolumn{2}{c}{ 3995.1 / 3766 = 1.06} \\ \hline
\end{tabular} \\
\flushleft{\footnotesize $^a$ Denotes fixed values. \newline
$^b$ Parameter unconstrained at 90\% within the model limits. \newline
$^c$ The reflection fraction in the lamppost model was calculated self-consistently, and is not itself a free parameter. \newline
Errors shown represent 90\% confidence intervals.}
\end{minipage}
\end{table*}

%%%%% TABLE 3: RelxillD Model v7 w/ Gauss P05 %%%%%
\begin{table*}
\begin{minipage}{2\columnwidth}
\caption{Co-Fit Spectral Results with Model 2 \label{tab:relxillDgauss}}
\centering
\renewcommand{\arraystretch}{1.25}
\begin{tabular}{lcccc}
\multicolumn{5}{c}{\textsc{constant$^\ast$tbabs$^\ast$(diskbb + relxillD + gaussian)}} \\
\hline \hline
Model & Parameter & Units & Low Flux Value & High Flux Value \\ \hline
\textsc{constant} & --- & FPMA/B &  \multicolumn{2}{c}{ $0.981 \pm0.002$ } \\ \hline
\multirow{2}{*}{\textsc{tbabs}}
  & $N_{\rm H}$ (A) & $10^{21}$\,cm$^{-2}$ & \multicolumn{2}{c}{ 3.2$^{~a}$ } \\
  & $N_{\rm H}$ (B) & $10^{21}$\,cm$^{-2}$ & \multicolumn{2}{c}{ $6.3 \pm0.4$ } \\ \hline
\multirow{2}{*}{\textsc{diskbb}}
  & $T_{\rm in}$ & keV & $1.060 \pm0.003$ & $1.129 \pm0.004$ \\
  & $N$ & --- & $456^{+6}_{-5}$ & $314^{+4}_{-5}$ \\ \hline
\multirow{10}{*}{\textsc{relxillD}}
  & $q_1$ & --- & \multicolumn{2}{c}{ $> 9.1^{~b}$ } \\
  & $q_2$ & --- & \multicolumn{2}{c}{ $3.0^{~a}$ }  \\
  & $R_{\rm break}$ & $GM/c^2$ & \multicolumn{2}{c}{ $> 7.5^{~b}$ } \\
  & $a$ & --- & \multicolumn{2}{c}{ $0.988^{+0.004}_{-0.010}$ } \\
  & Incl. & deg. & \multicolumn{2}{c}{ $75 \pm2$ } \\
  & $R_{\rm in}$ & ISCOs & \multicolumn{2}{c}{ 1.0$^{~a}$ } \\
  & $R_{\rm out}$ & $GM/c^2$ & \multicolumn{2}{c}{ 1000$^{~a}$ } \\
  & $\Gamma$ & --- & $2.35 \pm0.01$ & $2.35 \pm0.01$ \\
  & log\,$\xi$ & log\,(erg\,cm\,s$^{-1}$) & $3.70 \pm0.03$ & $3.94 \pm0.03$ \\
  & $A_{\rm Fe}$ & solar & \multicolumn{2}{c}{ $1.0^{~a}$ } \\
  & log\,$n_e$ & log\,(cm$^{-3}$) & \multicolumn{2}{c}{ $> 18.98^{~b}$ } \\
  & $f_{refl}$ & --- & $3.0 \pm0.2$ & $4.5^{+4.3}_{-0.8}$ \\
  & $N$ & $10^{-2}$ & $1.54\pm0.04$ & $1.62^{+0.91}_{-0.04}$ \\ \hline
  & $E_{\rm line}$ & keV & \multicolumn{2}{c}{ $6.87 \pm0.07$ } \\
\textsc{gaussian} & $\sigma_{\rm line}$ & eV & \multicolumn{2}{c}{ 50$^{~a}$ } \\
  & $N_{\rm line}$ & $10^{-4}$\,ph\,cm$^{-2}$\,s$^{-1}$ & \multicolumn{2}{c}{ $-2.6 \pm0.9$ } \\ \hline
  & $\chi^{2}$/dof & --- & \multicolumn{2}{c}{ 3998.7 / 3765 = 1.06 } \\ \hline
\end{tabular}
\flushleft{\footnotesize $^a$ Denotes fixed values. \newline
$^b$ Parameter consistent with hard limit. \newline
Errors shown represent 90\% confidence intervals.}
\end{minipage}
\end{table*}

%%%%% FIGURES 6 + 7, Spectra %%%%%
\begin{figure*}
\begin{center}
    \includegraphics[width=8.5cm,trim=1.5cm 1.3cm 3.5cm 1.8cm, clip]{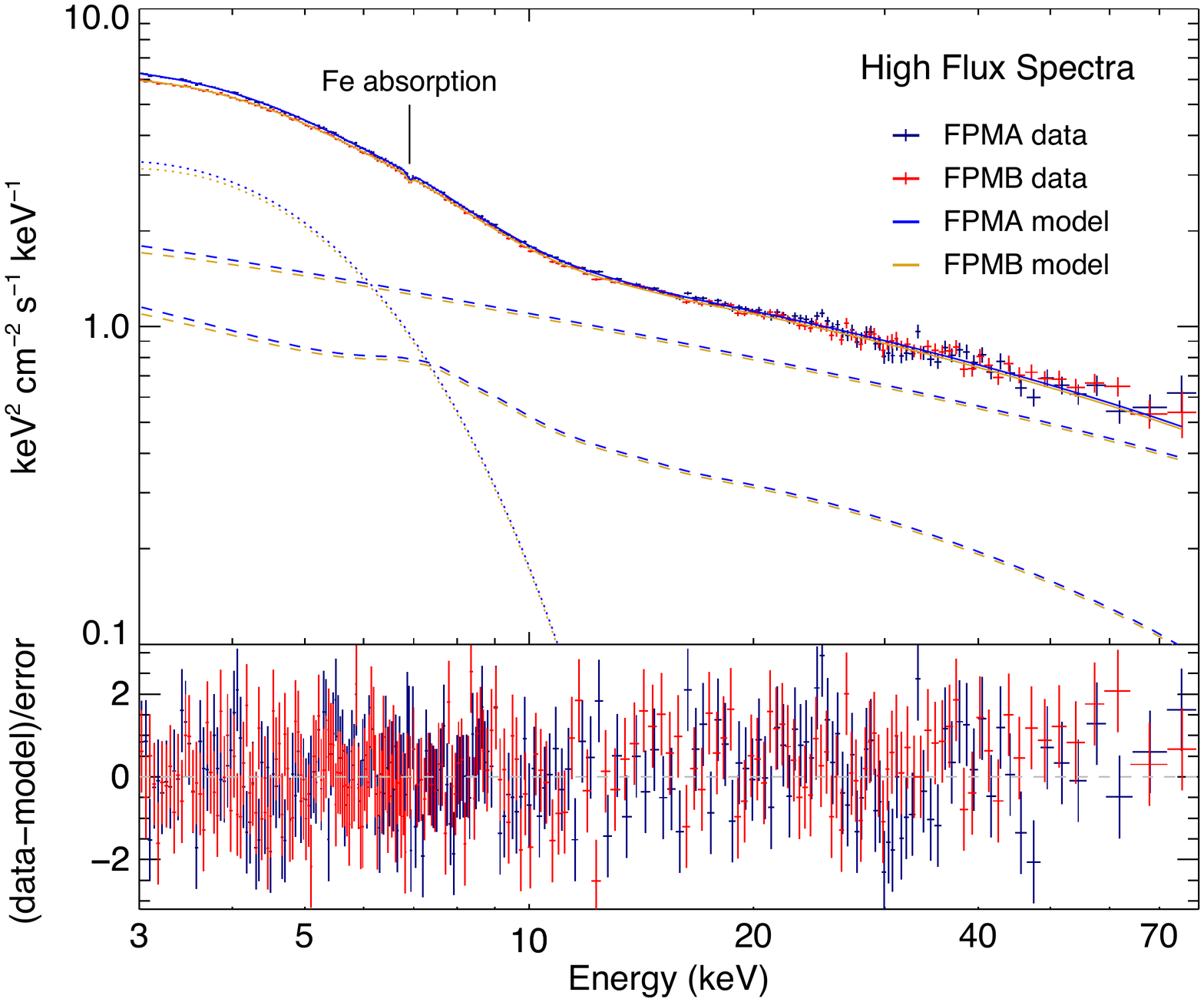}
    \includegraphics[width=8.5cm,trim=1.5cm 1.3cm 3.5cm 1.8cm, clip]{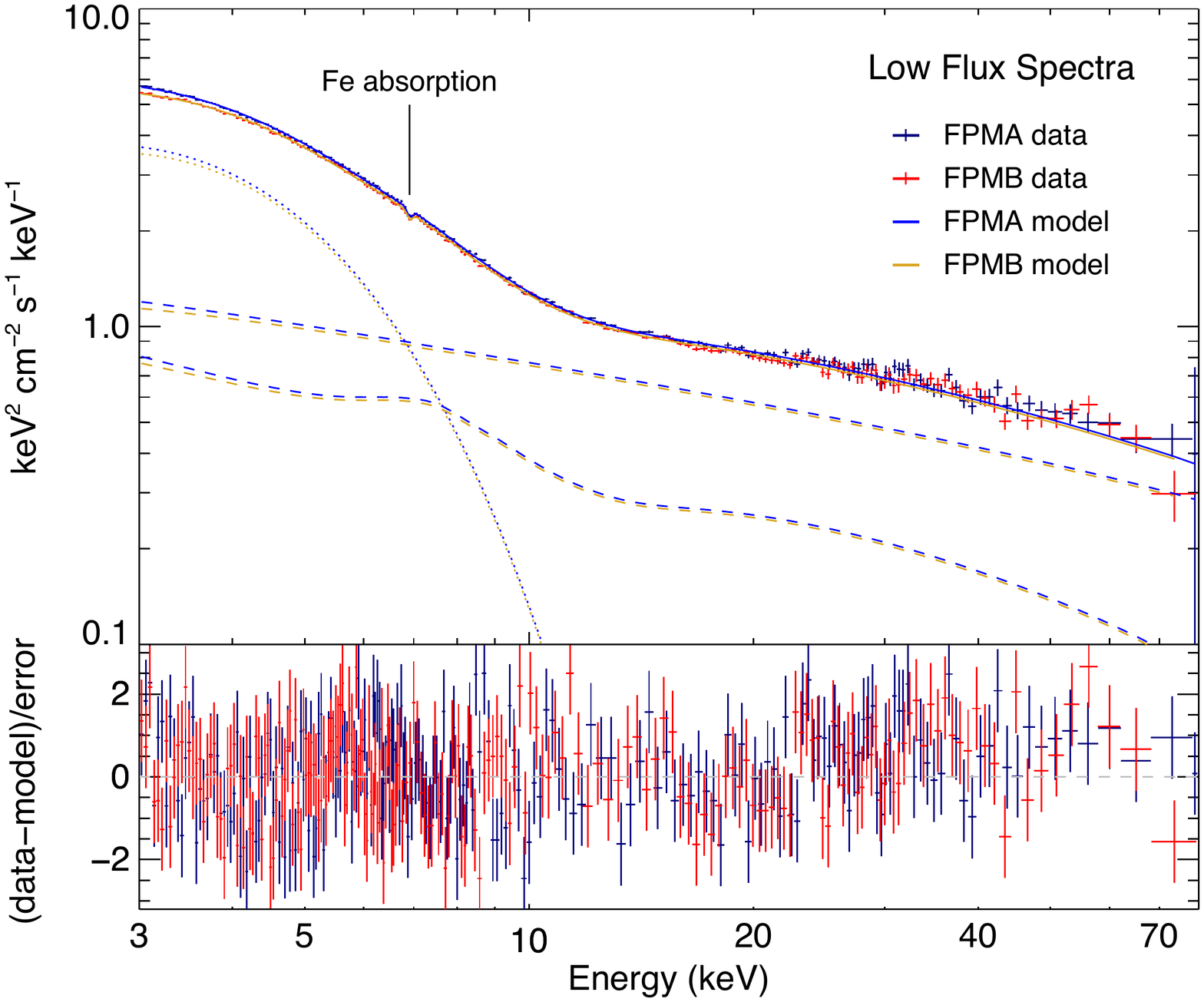}
    \caption{Model~1 unfolded spectra and normalized residuals for the high flux (left) and low flux (right) spectra, calculating using a high density lamppost geometry and a narrow absorption component (see Table~\ref{tab:relxillLpDgauss}).}
\label{fig:spectra1}
\end{center}
\end{figure*}

\begin{figure*}
\begin{center}
    \includegraphics[width=8.5cm,trim=1.5cm 1.3cm 3.5cm 1.8cm, clip]{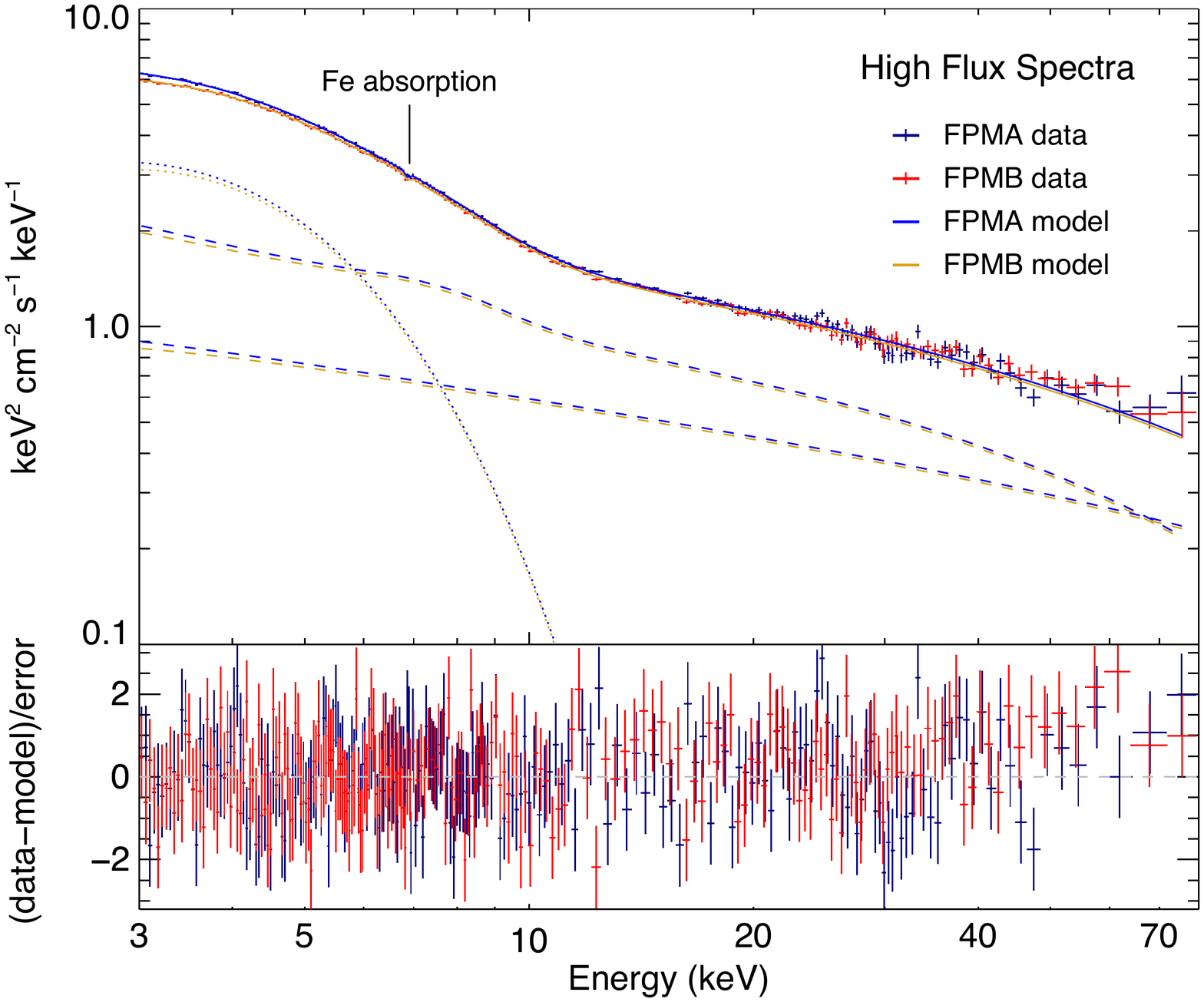}
    \includegraphics[width=8.5cm,trim=1.5cm 1.3cm 3.5cm 1.8cm, clip]{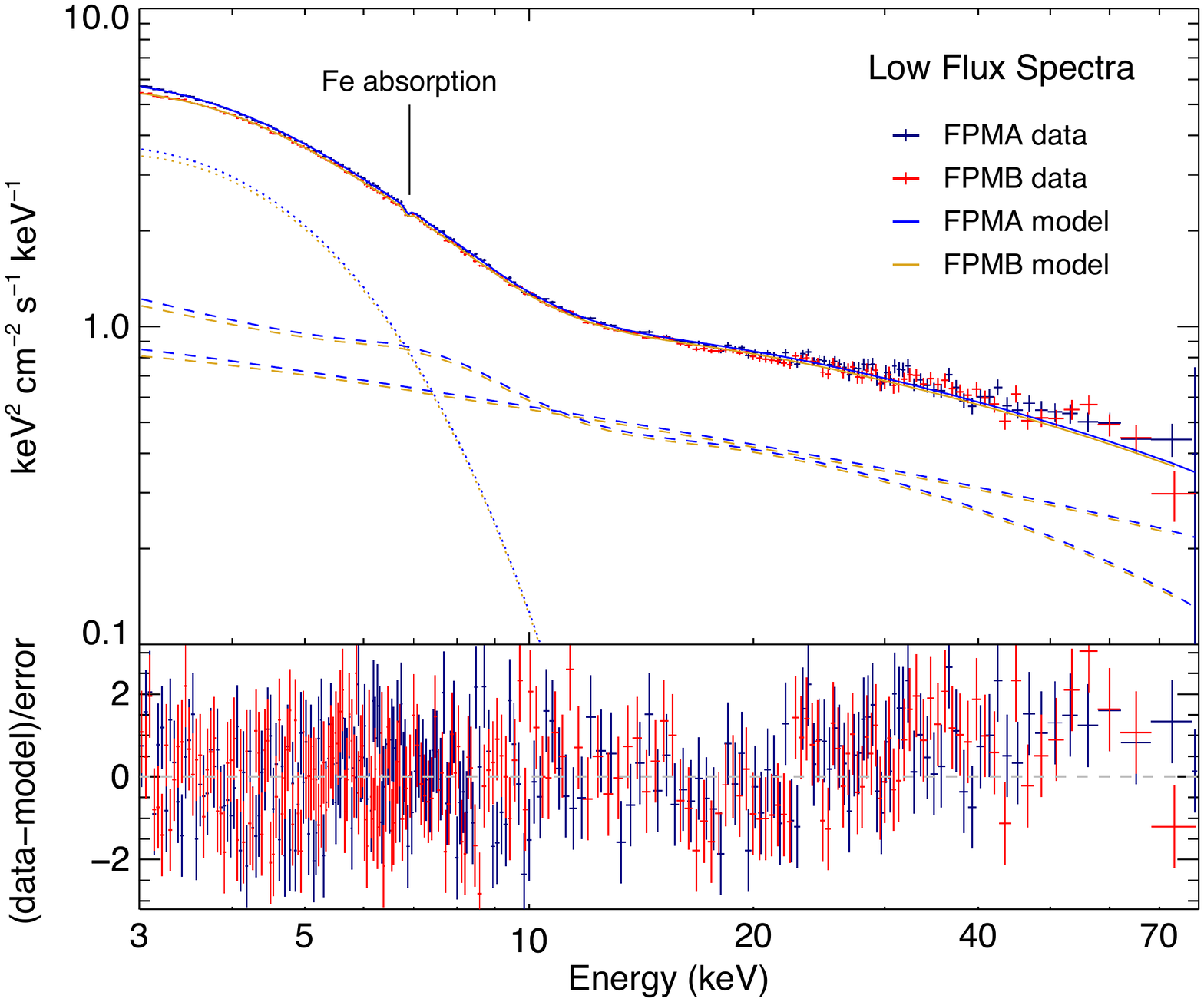}
    \caption{Model~2 unfolded spectra and normalized residuals for the high flux (left) and low flux (right) spectra, using \textsc{relxillD} with a broken power law emissivity and a narrow absorption component. Best-fit parameters are given in Table~\ref{tab:relxillDgauss}.}
\label{fig:spectra2}
\end{center}
\end{figure*}

In the residuals to both of our high density reflection models, there appears a possible absorption feature just below 7\,keV. Absorption lines may represent the presence of a disk wind, which had been previously reported in MAXI~J1803 \citep{miller21, matasanchez22}. We therefore tested for the presence of a narrow absorption line by adding a Gaussian component to both models, with a negative normalization and with the Gaussian width fixed to $\sigma=50$\,eV, which we choose to represent a narrow line as it is much smaller than the 400\,eV energy resolution of NuSTAR. The inclusion of the additional component improved both fits, with $\Delta \chi^{2} =  29.0$ and 12.6 for the \textsc{relxillLpD} and \textsc{relxillD} models, respectively. These results are provided in Tables~\ref{tab:relxillLpDgauss} and \ref{tab:relxillDgauss}, and hereafter we refer to these models respectively as Model~1 and Model~2, for the sake of brevity. We measure a line energy of $6.91 \pm0.06$\,keV (Model~1) and $6.87 \pm0.07$\,keV (Model~2), so the line energy is consistent between the two models. The line normalizations are also consistent within errors, at $-3.6\pm1.1 \times10^{-4}$\,ph\,cm$^{-2}$\,s$^{-1}$ and $-2.6\pm0.9 \times10^{-4}$\,ph\,cm$^{-2}$\,s$^{-1}$, for Models~1 and 2, respectively. To determine the significance of the line, we ran a \textsc{steppar} calculation in XSPEC over a range of values for the Gaussian normalization. A zero normalization is ruled out at the 5\,$\sigma$ level in the case of the lamppost model, and at almost $5\,\sigma$ in the case of the standard high density model.

As an additional test, we simulated 5000 continuum spectra (with no absorption line) using the \textsc{simftest} procedure in XSPEC to check whether random statistical fluctuations might produce a feature as strong as that observed (searching from 5--10\,keV), with just 2 simulations (or 0.4\%) meeting that criteria. This corresponds to a $\sim 3.5\,\sigma$ significance for the absorption line component, and so we include the narrow absorption line in our spectral modeling. In either model, the addition of the line does not significantly impact the other spectral parameters.

The high and low flux spectra and model residuals for both models are plotted in Figures~\ref{fig:spectra1} and \ref{fig:spectra2}. Considering our results presented in Tables~\ref{tab:relxillLpDgauss} and \ref{tab:relxillDgauss}, it becomes clear that higher disk densities are important when fitting the MAXI~J1803 spectra. One limitation of the high density reflection models is a fixed high-energy cutoff for the incident power law at 300\,keV, which is responsible for the residuals $\gtrsim$\,30\,keV in the right lower panels of Figures~\ref{fig:spectra1} and \ref{fig:spectra2}, and yet higher disk densities still provide a significantly improved fit over models with a variable high-energy cutoff. Both models have a best-fit value pegged at the upper limit of $n_e = 10^{19}$\,cm$^{-3}$, and lower limits of log\,$n_e > 18.76$ (Model~1) and $> 18.98$ (Model~2). Both models are consistent with an accretion disk inclination of $\sim$75 degrees, and both models provide similar interpretations of the differences between the high and low flux spectra --- for both models, the higher flux spectra are fit with a higher disk continuum temperature, a lower disk continuum normalization, and a greater normalization for the non-thermal power law component. Likely a consequence of the increase in flux illuminating the accretion disk, the disk ionization also increases from log\,$\xi$ = $3.70 \pm0.03$ to $3.94 \pm0.03$\footnote{The ionization parameter is defined as $\xi = 4\pi F_X/n_e$, in units of erg\,cm\,s$^{-1}$, where $F_X$ is the incident X-ray flux.}. Despite the different continuum disk parameters between the high and low flux spectra, the shape of the disk spectrum is very similar between the two (this is true for both models, and can be seen in Figures~\ref{fig:spectra1} and \ref{fig:spectra2}). While both models agree on the changes between the high and low flux spectra, there are important differences between our results for each model. While Model~1 (in the lamppost geometry) cannot constrain the BH spin, Model~2 measures a rapidly spinning BH with $a$ = $0.988^{+0.004}_{-0.010}$. When allowing for a free Fe abundance, Model~1 measures $A_{\rm Fe}$ = $1.4^{+0.2}_{-0.4}$ times the solar abundance, while Model~2 cannot constrain $A_{\rm Fe}$ and we leave it fixed to the solar value. In the lamppost geometry, our results from Model~1 suggest a distant corona; we measure a lamppost height of $h = 31^{+8}_{-7}\,R_g$ for the low flux spectra and $h = 45^{+18}_{-15}\,R_g$ for the high flux spectra. Model~2 instead relies on a broken power law disk emissivity, with $R_{\rm break} > 7.5\,R_g$, $q_{1} > 9.1$, and $q_{2}$ fixed to the non-relativistic limit of 3.

\begin{figure*}
\begin{center}
    \includegraphics[width=8.5cm,trim=4.5cm 3.8cm 6cm 5cm, clip]{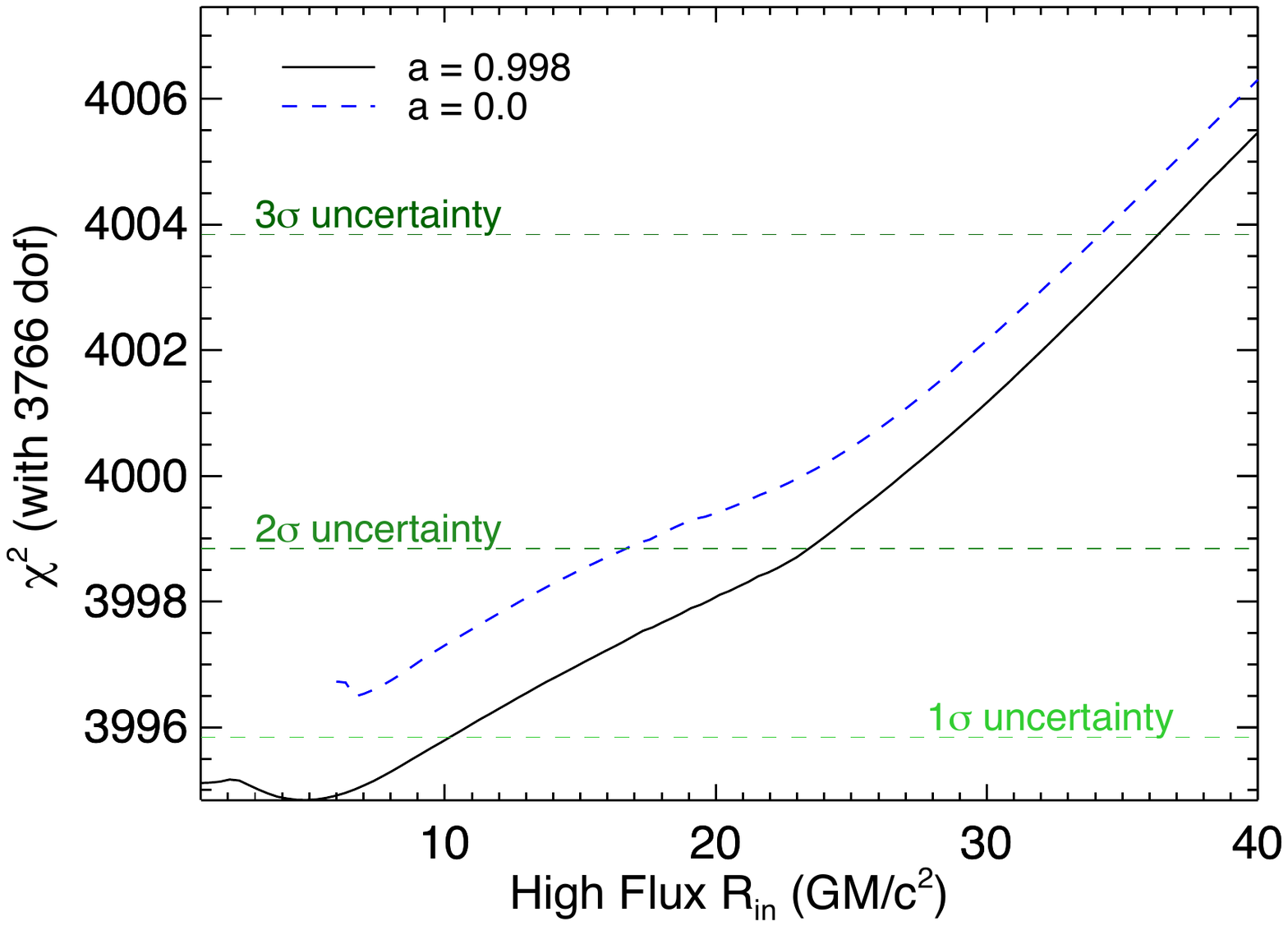}
    \includegraphics[width=8.5cm,trim=4.5cm 3.8cm 6cm 5cm,, clip]{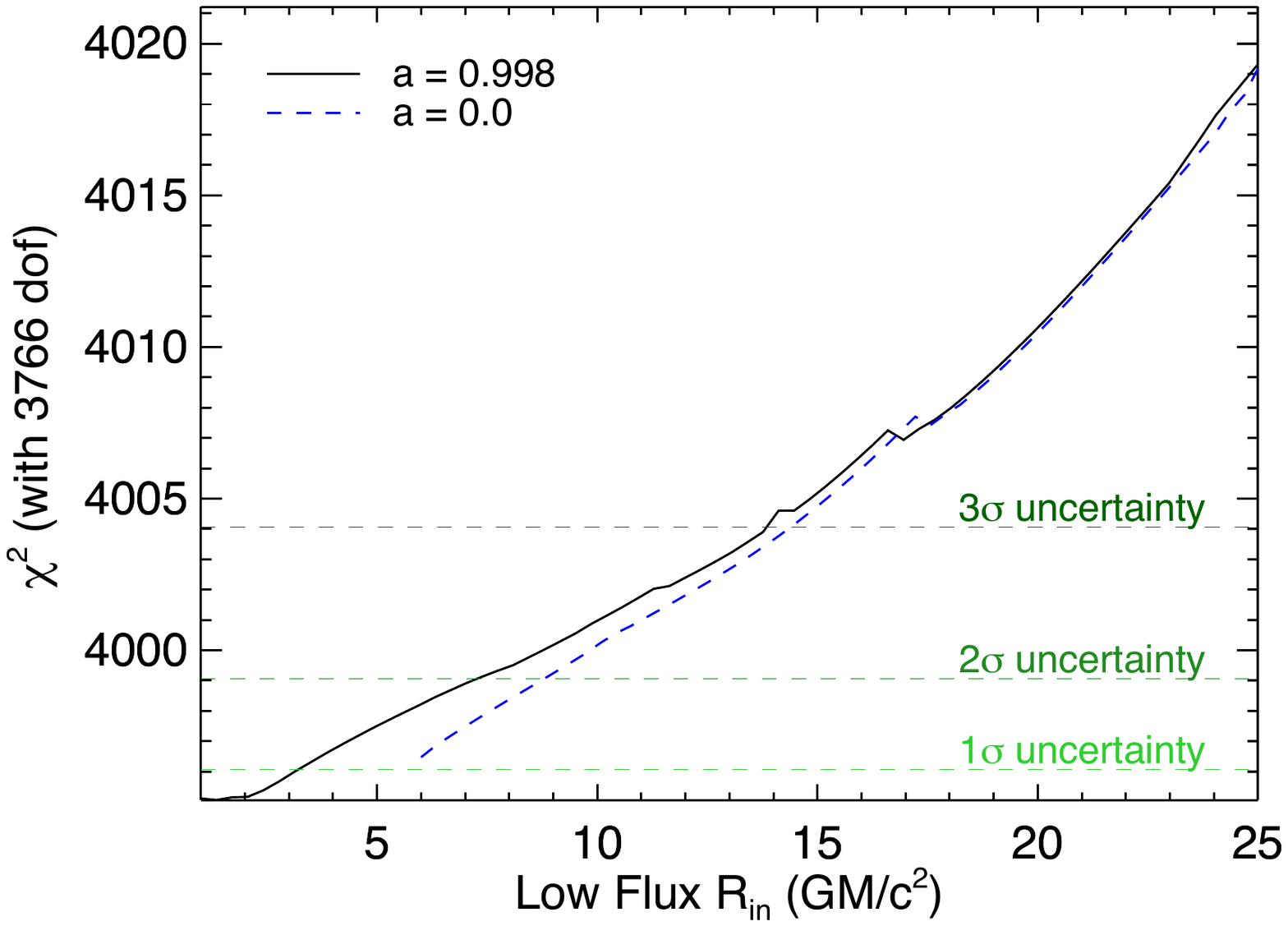}
    \caption{Contour plots for the inner disk radius from the best-fit high density lamppost model (Model 1). The inner disk radius was allowed to vary separately for the high flux (left) and low flux (right) spectra while co-fitting the spectra. Contours were produced for a maximally rotating BH (black, solid line) as well as for a non-rotating BH (blue, dashed line).}
\label{fig:diskradius}
\end{center}
\end{figure*}

Considering that we are unable to measure the BH spin in the lamppost geometry, we also tested for the possibility of a truncated disk in our observation of MAXI~J1803 using the reflection spectrum. For both models, we fixed the spin to represent either a maximally spinning or non-rotating BH and allowed for the inner disk radius ($R_{\rm in}$) to be free. With Model~1, we are able to rule out a highly truncated accretion disk, with contours favoring a disk that approaches the ISCO and with a $90\%$ limit of $R_{\rm in} < 20\,R_g$, as shown in Figure~\ref{fig:diskradius}. Model~2, on the other hand, significantly disfavors any truncation of the accretion disk.

%%%%% FIGURE 8, R_IN %%%%%

\section{Discussion} \label{sec:discuss}

We observed MAXI~J1803 beginning on May 14, resulting in over 30\,ks of exposure time in each NuSTAR FPM, near the peak of the outburst. During our observation, the source showed variability on the timescale of an hour, and we divided the observation into low and high flux spectra to characterize this variability and discern any changes in the accretion geometry that may be responsible for the observed changes in flux. We also investigated variability on shorter timescales for both the high and low flux segments of the observation.

On shorter timescales, the low flux PDS showed two potential QPO signals at 5.4\,Hz and 9.4\,Hz. Since these two frequencies are not harmonically related, these signals must be entirely separate. Both signals are statistically significant, but with lower quality factors than expected for a strong QPO signal. In MAXI~J1803, a Type~C QPO was first detected by NICER at 0.13\,Hz, increasing to 0.26\,Hz by May 4 \citep{bult21}. This QPO evolved as the source continued to rise in the hard state, to 0.41\,Hz as seen by NuSTAR on May 5 \citep{xu21}, and when the source transitioned into a hard-intermediate state the QPO was seen to evolve from 5.31\,Hz to 7.61\,Hz with AstroSat \citep{jana22}. We therefore suggest that the higher of the possible QPOs we detect is Type~C, though without the flat-top broadband noise that was seen in the AstroSat data and that typically accompanies Type~C QPOs \citep[see][for a recent review]{ingram19}. This is due to another state transition --- our NuSTAR observation beginning on May 14 catches MAXI~J1803 in the soft-intermediate state, and the source is much less variable on sub-second timescales when compared to the preceding AstroSat observation.

We also suggest that the potential lower frequency QPO (at 5.4\,Hz) is Type~B, detected in our NuSTAR observation as MAXI~J1803 was in the soft-intermediate state. Unfortunately, NICER was unable to cover MAXI~J1803 from May 5 until May 18. With 1.4\,ks of exposure on May 18, NICER was able to detect a Type~B QPO at 6\,Hz along with a subharmonic at 3\,Hz \citep{ubach21}. We suggest this is the same QPO we detect at 5.4\,Hz, which means this is the first time NuSTAR has ever observed a Type~B QPO, or two distinct QPOs at the same time. Type~B and C QPOs were also detected simultaneously without any flat-top broadband noise in RXTE observations of GRO~1655$-$40, when that source was in the ultra-luminous state at the peak of its 2005 outburst \cite{motta12}. We observe MAXI~J1803 near the peak of its outburst (see Fig.~\ref{fig:lchc}), yet the most puzzling aspect of the potential QPOs we detect is that they are only present in the low flux segments of the light curve. While the low flux PDS exhibits the potential QPOs, the high flux PDS is featureless apart from broadband noise. Considering that the potential QPO signals are very broad or weak, and that the overall rms variability in either case is below 10\%, we unfortunately cannot investigate these signals further with the available data.

In our spectral analysis, phenomenological fits showed clear signatures of relativistic reflection (Fig.~\ref{fig:feline}), and so we tested the full range of reflection models in the \textsc{relxill} family. Initial modeling with \textsc{relxill} and \textsc{relxillLp} struggled to adequately fit the MAXI~J1803 spectra, while providing poor constraints on the disk inclination and requiring high Fe abundances. Models which allow for disk densities higher than $n_e$ = 10$^{15}$\,cm$^{-3}$ provide a much better statistical fit to the data. Our high density model with the lamppost coronal geometry, Model~1, is also able to constrain the Fe abundance to be consistent with the solar value (see Table~\ref{tab:relxillLpDgauss}). When using the high density version of the standard \textsc{relxill} model (Model~2), our fits were significantly improved when allowing for higher densities, though the Fe abundance was still poorly constrained and would tend towards the upper limit of 10 times the solar value when left free to vary. In both cases, the density parameter was pegged at the model upper limit of log\,$n_e$ = 19. Our results with MAXI~J1803 provide yet another piece of evidence that the effects of higher disk densities are important when considering reflection studies of BH XRBs.

Despite ruling out a highly truncated disk (where $R_{in} \gtrsim$\,100 $R_g$) in the lamppost geometry (see Figure~\ref{fig:diskradius}), we are unable to constrain the BH spin when using Model~1. This is a result of the lamppost height for both the high and low flux spectra, which we measure to be $h = 45^{+18}_{-15}$ and $31^{+8}_{-7}$ $R_g$, respectively. At such a height, differences in the ISCO due to the BH spin produce a negligible difference in the relativistic blurring of the Fe line \citep[see Figure~9e of][]{dauser13}. \cite{fabian14} also show that a robust determination of the BH spin in the lamppost geometry depends on the corona being low enough ($h < 10$) to properly illuminate the ISCO. A similar result with lamppost models was seen by \cite{draghis20} when comparing multiple reflection models fit to the spectra of EXO~1846$-$031 during its 2019 outburst.

In contrast, when we do not impose the lamppost geometry on MAXI~J1803 and instead allow for a more flexible disk emissivity profile, our results with Model~2 rule out any disk truncation and allow for an estimate of the BH spin. We measure $a = 0.988^{+0.004}_{-0.010}$, and suggest that MAXI~J1803 contains a very rapidly spinning BH. This measurement is consistent with a different intermediate state NuSTAR observation of MAXI~J1803 on May 23, which, alongside NICER data, was best-fit with $a \sim 0.991$ \citep{feng22}, as well as a continuum measurement of the thermal disk spectrum using AstroSat data from May 11 which suggested $a > 0.7$ \citep{chand22}. We do advise some caution in interpreting our measurement of the BH spin with Model~2. While we quote the 90\% statistical error, systematic uncertainties when applying reflection models (including the assumptions and simplifications inherent in the models themselves) could be on the order of or greater than the statistical uncertainties. This question is considered by \cite{draghis20} in the case of EXO~1846$-$031, and the authors conclude that statistical and systematic uncertainties are comparable. While the two models we present are very similar in terms of their spectral shape and fit quality, as well as the fact that all shared parameters are either consistent or have overlapping confidence intervals, each model paints a very different picture of the accretion geometry of MAXI~J1803.

In addition to our reflection results, we also find strong evidence for a narrow absorption feature in the NuSTAR spectra. The line is more significantly detected in Model~1 at $6.91 \pm0.06$\,keV. Optical disk wind signatures were observed for MAXI~J1803 while in the hard state \citep{matasanchez22}, and in X-rays a disk wind was reported based on Swift/XRT observations from May 20, with an absorption line feature at 6.9\,keV and with an equivalent width of $\sim$30\,eV (\cite{miller21}, private correspondence). We measure an equivalent width of $5.5 \pm1.7$\,eV for the high flux spectra, and $7.1^{+2.2}_{-2.3}$\,eV for the low flux spectra. This suggests that the disk wind feature intensifies or broadens between our May 14 observation and that of \cite{miller21} on May 20. Assuming the line is due to Fe\,\textsc{xxv} absorption blueshifted from 6.70\,keV to the observed 6.91\,keV, this corresponds to a wind velocity of $0.03\,c$. The line could also occur due to Fe\,\textsc{xxvi} absorption with no shift, since the measured line energy is consistent with 6.97\,keV at the 90\% level.

If we consider the results of both our models and what they share in common, a conservative but clear description of MAXI~J1803 can be reached. First, even from our phenomenological modeling it was clear that the hard component (in our case, a power law) was responsible for the slow variability evident in the source light curve (Figure~\ref{fig:lchid}). This is true when considering our reflection modeling results, as the normalization of the incident power law component responsible for reflection is greater during periods of high flux, while the disk continuum flux contribution remains roughly the same throughout our observation. During periods of higher flux, the disk ionization also increased, which is to be expected by definition, since $\xi \propto$ $F_X$ (where $F_X$ is the incident flux in X-rays). In our observation, the long term variability is driven by changes in the corona rather than the accretion disk. Considering the source geometry, we can also be confident that the accretion disk is not truncated beyond $\sim 20\,R_g$, since this is the most conservative 90\% limit given by the lamppost configuration (Model~1, see Figure~\ref{fig:diskradius}). Finally, we find that when using high density reflection models, we are able to constrain the inclination of the inner accretion disk to $\sim75$ degrees. This falls within the expected orbital inclination of 60--80 degrees, considering MAXI~J1803 showed periodic absorption dips early in its outburst \citep{homan21, xu21, jana22}. Both models proposed here yield consistent results for this measurement.

\section{Conclusions} \label{sec:conclude}

We report on the results of a NuSTAR ToO of MAXI~J1803, finding strong evidence for a highly inclined ($\sim$75 degrees), high density accretion disk, as well an absorption feature at 6.9\,keV that may be due to an outflowing disk wind. We also report the detection of a potential Type~B QPO at 5.4\,Hz, as well as a possible Type~C QPO at 9.4\,Hz. On much longer timescales the coronal emission of the source is variable, which we investigate by fitting spectra extracted during periods of lower and higher flux separately. We successfully fit the spectra with two distinct models, each with its own interpretation, allowing either for an estimate of the BH spin or a constraint of the Fe abundance in the accretion disk. In either case, our findings agree with other investigations of MAXI~J1803 which suggest a near-maximally spinning BH in a highly inclined orbit with its companion, and outflowing disk winds throughout its 2021 May outburst.

\section{Acknowledgements}

BMC and JAT acknowledge partial support from NASA under Astrophysics Data Analysis Program (ADAP) grant 80NSSC19K0586 and under NuSTAR Guest Observer grant 80NSSC22K0059. JJ acknowledges support from the Leverhulme Trust, the Isaac Newton Trust and St Edward's College, University of Cambridge. JH acknowledges support from an appointment to the NASA Postdoctoral Program at the Goddard Space Flight Center, administered by Oak Ridge Associated Universities through a contract with NASA. 

We also thank Jon Miller for discussing the Swift/XRT detection of a disk wind in MAXI~J1803 \citep{miller21}, as well as Brian Grefenstette and Kristen Madsen for their time and suggestions concerning the calibration and agreement between NuSTAR's two FPMs.

\bibliography{refs_Coughenour}{}
\bibliographystyle{aasjournal}

%%% For this sample we use BibTeX plus aasjournals.bst to generate the
%%% the bibliography. The sample631.bib file was populated from ADS. To
%%% get the citations to show in the compiled file do the following:
%%%
%%% pdflatex sample631.tex
%%% bibtext sample631
%%% pdflatex sample631.tex
%%% pdflatex sample631.tex


\end{document}